\documentclass{aa}  
\usepackage{graphicx}
\usepackage{txfonts}
\def\mrk{Mrk~509}
\def\mrkt{Mrk~279}

\def\ch{{\it Chandra}}

\def\inte{{\it INTEGRAL}}
\def\epi{EPIC-PN}

\def\xmm{XMM-{\it Newton}}

\def\kms{km\,s$^{-1}$}
\def\Halpha{\ifmmode {\rm H}\alpha \else H$\alpha$\fi}
\def\Hbeta{\ifmmode {\rm H}\beta \else H$\beta$\fi}
\def\Hgamma{\ifmmode {\rm H}\gamma \else H$\gamma$\fi}
\def\Hdelta{\ifmmode {\rm H}\delta \else H$\delta$\fi}
\def\Lya{\ifmmode {\rm Ly}\alpha \else Ly$\alpha$\fi}
\def\Lyb{\ifmmode {\rm Ly}\beta \else Ly$\beta$\fi}
\def\Lyg{\ifmmode {\rm Ly}\beta \else Ly$\gamma$\fi}
\def\fek{{Fe\,K$\alpha$}}
\def\fekb{{Fe\,K$\beta$}}

\def\heii{He\,{\sc ii}}

\def\ciii{\ifmmode {\rm C}\,{\sc iii} \else C\,{\sc iii}\fi}
\def\civ{\ifmmode {\rm C}\,{\sc iv} \else C\,{\sc iv}\fi}
\def\cv{\ifmmode {\rm C}\,{\sc v} \else C\,{\sc v}\fi}
\def\cvi{\ifmmode {\rm C}\,{\sc vi} \else C\,{\sc vi}\fi}

\def\niii{N\,{\sc iii}}

\def\nv{N\,{\sc v}}

\def\nvii{N\,{\sc vii}}
\def\oi{O\,{\sc i}}

\def\o5007{[O\,{\sc iii}]\,$\lambda5007$}

\def\ovi{O\,{\sc vi}}
\def\ovii{O\,{\sc vii}}
\def\oviii{O\,{\sc viii}}

\def\neix{Ne\,{\sc ix}}

\def\siiv{Si\,{\sc iv}}

\def\gtsim{\raisebox{-.5ex}{$\;\stackrel{>}{\sim}\;$}}
\begin{document}
\title{Multiwavelength campaign on Mrk 509}

\subtitle{XV. A global modeling of the broad emission lines in the Optical, UV and X-ray bands}

 \author{E.~Costantini, 
          \inst{1,2}
	G.~Kriss, 
	\inst{3}
	J.S.~Kaastra, 
	\inst{1,4}
	S.~Bianchi,
	\inst{5}
	G.~Branduardi-Raymont,
	\inst{6}
	M.~Cappi,
	\inst{7}
	B. De Marco,
	\inst{8}
	J.~Ebrero,
	\inst{9}
	M.~Mehdipour,
	\inst{1}
	P.-O.~Petrucci,
	\inst{10,11}
	S.~Paltani,
	\inst{12}
	G.~Ponti,
	\inst{8}
	K.C.~Steenbrugge
	\inst{13}
	\and
	N.~Arav
	\inst{14}
}
		
\institute{\inst{1}SRON, Netherlands Institute for Space Research, Sorbonnelaan, 2, 3584, CA, Utrecht, The Netherlands\\ 
              \email{e.costantini@sron.nl}\\        
    	\inst{2}Anton Pannekoek Institute, University of Amsterdam, Postbus 94249, 1090 GE Amsterdam, The Netherlands\\ 	     
	\inst{3}Space Telescope Science Institute, 3700 San Martin Drive, Baltimore, MD, 21218, USA\\ 		 
	\inst{4}Leiden Observatory, Leiden University, PO Box 9513 2300 RA Leiden, the Netherlands\\ 	 
	\inst{5}Dipartimento di Matematica e Fisica, Universit\'{a} degli Studi Roma Tre, via della Vasca Navale 84, 00146 Roma, Italy\\ 	 
	\inst{6}Mullard Space Science Laboratory, University College London, Holmbury St. Mary, Dorking, Surrey, RH5 6NT, UK\\ 	 
	\inst{7}INAF-IASF Bologna, via Gobetti 101, 40129 Bologna, Italy\\ 	 
	\inst{8}Max-Planck Instit\"{u}t f\"{u}r extraterrestrische Physik, Giessenbachstrasse 1, D-85748, Garching bei M\"{u}nchen, Germany\\ 
	\inst{9}XMM-Newton Science Operations Centre, ESAC, P.O. Box 78, E-28691 Villanueva de la Ca\~{n}ada, Madrid, Spain\\ 	 
	\inst{10}Univ. Grenoble Alpes, IPAG, F-38000 Grenoble, France\\
	\inst{11}CNRS, IPAG, F-38000 Grenoble, France\\ 
	\inst{12}ISDC Data Centre for Astrophysics, Astronomical Observatory of the University of Geneva, 16 ch. d'Ecogia, 1290 Versoix, Switzerland\\ 
	 \inst{13}Instituto de Astronom\'{i}a, Universidad Cat\'{o}lica del Norte, Avenida Angamos 0610, Antofagasta, Chile\\ 
	\inst{14}Department of Physics, Virginia Tech, Blacksburg, VA 24061, USA\\
}
\date{Received/accepted/}

\abstract{}{We model the broad emission lines present in the optical, UV and X-ray spectra of Mrk~509,  
 a bright type 1 Seyfert galaxy. The broad lines were simultaneously observed during a large multiwavelength campaign, using the 
 \xmm-OM for the optical lines, HST-COS for the UV lines and \xmm-RGS and Epic for the X-ray lines respectively. We also used FUSE 
 archival data for the broad lines observed in the far-ultra-violet. The goal is to find a physical connection among the lines 
 measured at different wavelengths and determine the size and the distance from the central source of the emitting gas components.}{ 
 We used the "Locally optimally emission Cloud" (LOC) model which interprets the emissivity of the broad line region (BLR) as regulated by 
 powerlaw distributions of both gas density and distances from the central source.}{ We find that one LOC component cannot model all the lines simultaneously. In particular, we find that
the X-ray and UV lines likely may originate in the more internal part of the AGN, at radii 
 in the range $\sim5\times10^{14}-3\times10^{17}$\,cm, while the optical lines and part of the UV lines may likely be originating further out, at radii 
 $\sim3\times10^{17}-3\times10^{18}$\,cm.  These two gas components are parametrized by a radial distribution of the luminosities with a slope 
 $\gamma$ of $\sim1.15$ and $\sim1.10$, respectively, both of them covering at least 60\% of the source. 
 This simple parameterization points to a structured broad line region, with the higher ionized emission coming from closer in, while the 
 emission of the low-ionization lines is more concentrated in the outskirts of the broad line region.}{}


\authorrunning{E.~Costantini et al.}
\titlerunning{The broad line region of Mrk 509}

\maketitle
%

\section{Introduction}\label{par:intro}
The optical-UV spectra of Seyfert~1 galaxies and quasars is characterized by strong and broad emission lines, which are believed to be produced by gas
photoionized by the central source. The broad line region (BLR) gas has been initially proposed to be in the form of a set of clouds 
\citep[e.g.][]{kmt81}. However, first the confinement of these clouds was problematic \citep{mf87} and second, an accurate analysis of the smoothness 
of the broad line wings revealed that in fact the gas could not be in the form of discrete clouds but rather a continuous distribution of gas \citep{arav98}. Another hypothesis is that the gas is
ejected
from the accretion disc outskirts in the form of a wind \citep[e.g.][]{murray95,bottorff97,elvis00,czerny11}. Finally the gas reservoir could be provided by disrupted
stars in the vicinity of the black hole \citep{baldwin03}.\\
Extensive observation of this phenomenon, through the reverberation mapping technique, led to the conclusion that the BLR is extended over a large area 
and that the radius of the BLR scales with the square root of the ionizing luminosity \citep{peterson93}. 
The BLR does not have an homogeneous, isotropic distribution \citep[e.g.][]{decarli08}. Several studies pointed out that higher ionized lines
(represented by \ion{C}{iv}) are incompatible with an origin in the same region where the bulk of H$\beta$ is produced \citep{sulentic00}. Different approaches to this problem lead to
divergent results. A flat disk structure would preferentially emit \civ\, while H$\beta$ would be produced in a vertically extended region, more distant from the central source \citep{sulentic00,
decarli08,goad12}. Other interpretations propose a different scenario, where the gas emitting H$\beta$ has a flat geometry, near to the accretion disk, while \civ\ would be emitted in an
extended region \citep[][and references therein]{kz13}. By virtue of the tight correlation found between the BLR size and the AGN luminosity \citep{bentz13}, the BLR gas could also
arise from the accretion disk itself and generate a failed wind. The confinement of such cloud motion, involving both outflow and inflow of gas, would be set by the dust sublimation radius \citep{czerny11,galianni13}.  Studies of gas dynamics within the BLR point indeed to a complex motion of the gas \citep{pancoast12}, 
where the matter may sometimes infall towards the black hole \citep{pancoast13,gg13}. Observationally, the broad lines centroids often show shifts (up to hundreds \kms) 
with respect to the systemic velocity of the AGN. Higher-ionization lines (like \civ$\lambda1548$) show in general 
more pronounced blue-shifts with respect to lower ionization lines \citep{peterson97}. This points also to a stratified medium, where the illumination of the cloud is related to the
ionization of the clouds \citep{peterson04}. A way to model the BLR emission without a priori assumptions on its origin is the "locally optimally emitting cloud" model \citep[LOC, ][]{baldwin}, which describes the total emission of a line as a function of the density and
the distance of the gas from the central source \citep[see][ for a review]{korista97a}. This model has been successfully applied to the broad lines detected in the UV of e.g. \object{NGC~5548}
\citep[e.g.][]{korista00}. 

Emission from the BLR can in principle extend from the optical-UV up to the X-ray band. With the advent of \xmm\ and \ch,
relatively weak, but significant, broad emission lines have been detected in the soft X-ray band. Often these lines display a symmetric profile, suggesting an origin far from
the accretion disk where relativistic effects would instead distort the line profile \citep[e.g.][]{steen09,fabian09}. The most prominent X-ray lines with a non-relativistic
profile are found at the energy of the
\ovii\ triplet and the \oviii\,Ly$\alpha$ \citep[e.g.][ hereinafter C07]{boller07,steen09,longinotti10,ponti10,costantini07}. 

An extension of the LOC model, adding also the X-ray band  in the modeling, has been applied to
\object{Mrk~279} (C07). In that case, the luminosities of the soft-X-ray emission lines (\cvi, \nvii, \ovii, \oviii\ and \neix) were well predicted by the LOC model, suggesting also that the bulk
of the X-ray lines could possibly arise up to three times closer to the black hole than the UV lines.\\
A contribution of the BLR to the \fek\ line at 6.4\,keV has been often
debated. A comparison between the Full Width Half Maximum (FWHM) of the H$\beta$ line at 4861\,\AA\ and the FWHM of the narrow component of the \fek\ line as measured by \ch-HETG, did
not reveal any correlation, as it would have been expected if the lines originated from the same gas \citep{nandra06}. However, on a specific source, namely the liner \object{NGC~7213},  where no hard X-ray reflection was
observed, the \fek\ line and the \Hbeta\ line are consistent with having the same FWHM \citep{bianchi08}. On the other hand, as seen above, X-ray lines
 may originate in different regions of the BLR. Therefore a direct comparison between the FWHM of \fek\ and  H$\beta$ may not prove or disprove that \fek\ is also produced
in the BLR. A further extension of the LOC model to the 6.4\,keV region showed that, in the case of \mrkt, the BLR emission contributed for at most 17\% to the total \fek\ emission, suggesting that
reflection either from the disk or from the torus had to be instead the dominant emitter of that line \citep{costantini10}.

\object{Mrk~509} has been subject to a large multiwavelength campaign, carried out in
2009 \citep{kaastra1}. The source has been an ideal laboratory in order to study the ionized gas outflowing from the source \citep{detmers11,ebrero11,kaastra2,kriss11,steen11,arav12}. 
The broad band continuum was investigated in \citet[][]{med11,pop13,boissay14} and the \fek\ long term variability in \citet{ponti13}.
In this paper of the series we investigate the BLR emission through the emission lines, simultaneously detected by different instruments from the optical to the X-rays. 

The paper is organized as follows: In Sect.~\ref{par:data} the data are described. In Sect.~\ref{par:model} we describe the application of the LOC model to the data. The discussion is in
Sect.~\ref{par:discussion}, followed by the conclusions in Sect.~\ref{par:conclusion}. 

Here we adopt a redshift of 0.034397 \citep[][]{huchra93}. The cosmological parameters used are: 
H$_0$=70 km/s/Mpc, $\Omega_{\rm m}$=0.3, and $\Omega_{\Lambda}$=0.7. The errors are calculated at 1$\sigma$ significance, obtained using the $\chi^2$
statistical method.

\begin{table*}
\caption{\label{t:lines} Main parameters of the broad lines components used in this analysis.}
\begin{center}
\begin{tabular}{llllll}
\hline\hline
Ion & Wavelength & Lum$_{i}$ & Lum$_{b}$& Lum$_{vb}$& Inst.\\  
\hline
\fek & 1.93 & $-$& $-$ &$4.1\pm0.5$& 1\\
\neix &	13.69 &$-$ &$1.07\pm0.12$ &$-$&2\\
\oviii$^a$ &	18.96 &$-$&$1.56\pm0.14$&$-$ &2\\
\ovii&	22.1 &$-$ &$3.72\pm0.33$&$-$& 2\\
\nvii& 24.77 &$-$&$2.28\pm1.52$&$-$ &2\\
\cvi& 33.73 &$-$&$1.37\pm0.54$&$-$& 2\\
\hline
\ciii& 977& $-$  &       $39\pm      15$   &  $-$        &  3\\
\niii&991&		$-$    &    $-$     &   $35\pm     14$    		&  3\\
\ovi$^a$ & 1025&$-$   &      $62\pm      24$   &  $90\pm      36$&  3\\
\Lya$^a$&1216&$35\pm     7$  &   $ 196\pm      40$  &   $402\pm      80$&  4\\
\ion{N}{v} &1238&$-$  &      $ 15\pm     3 $  &   $-$   &  4\\
\ion{Si}{ii} &1260&$-$   &     $ 30\pm      6 $   &       $-$&  4\\
\ion{O}{i}$^a$ &1304&$-$   &      $14\pm     3 $  &     $-$&  4\\
\ion{C}{ii} &1335&$-$   &       $3.5\pm0.7$   &     $-$&  4\\
\ion{Si}{iv}$^a$ &1403&$-$     &    $55\pm    11 $  &   $-$    &  4\\
\ion{N}{iv}]&1486&$-$   &    $ 3.0\pm     0.6$   &   $-$    &  4\\
\ion{Si}{ii} & 1526&$-$  &      $ 6\pm   1 $  &   $-$   &  4\\
\ion{C}{iv} &1548&$49\pm      9 $ &    $124\pm      24$   &   $191\pm      39$&  4\\
\ion{He}{ii} &1640&$8.5\pm      1.7 $ &    $55\pm      11 $  &   $-$    &    4\\
\ion{O}{iii}] &1663&$-$   &     $27\pm      5$    &  $-$    &    4\\
\hline
\Hdelta & 4102 & $-$& $11\pm1$&$-$ & 5\\
\Hgamma$^a$& 4340 &$-$&$24\pm4$&$-$ & 5\\
\Hbeta & 4861 &$-$&$45\pm14$&$-$ & 5\\
\Halpha & 6563 &$-$&$121\pm9$&$-$ & 5\\
\hline
\end{tabular}
\end{center}
Notes:\\ 
In columns 3, 4 and 5, the lines luminosity are reported for the intermediate {\it(i)} with FWHM=1000--3000\,\kms, broad {\it(b)} with FWHM=4000--5000\,\kms, 
and very broad {\it(vb)} with FWHM$>$9000\,\kms\ components (defined in Sect.~\ref{par:uvlines}).\\
Restframe nominal wavelengths are in \AA, luminosities are in units of $10^{41}$\,erg\,s$^{-1}$.\\
Instruments: 1: \xmm-\epi, 2: \xmm-RGS, 3: FUSE, 4: HST-COS, 5: \xmm-OM\\
$^a$ Blends of lines: \oviii\ with the \ovii-He$\beta$ line; the \ovi\ doublet with \Lyb; The \Lya\ with the \ion{O}{v}] triplet and \heii; \oi\ with 
\ion{Si}{ii}; the \ion{Si}{iv} doublet with both the \ion{O}{iv}] and \ion{S}{iv} quintuplets; \Hgamma\ with \heii. 
\end{table*}

\section{The data}\label{par:data}
Here we make use of the analyses already presented in other papers of this series on \mrk. In particular the \xmm-OM optical lines are taken from \citet{med11}, the COS and FUSE broad
emission line fluxes are taken from \citet[][hereinafter K11]{kriss11}. The X-ray broad line data are from \citet[][using RGS and LETGS ]{detmers11, ebrero11} and  \citet[][using PN]{ponti13}. 

The lines that we use in our modeling are listed in Table~\ref{t:lines}.

\subsection{The X-ray broad lines}
The \xmm-RGS spectrum shows evidence of broad emission at energies consistent with the transitions of the main He-like 
(\ion{O}{vii} and \ion{Ne}{ix} triplets) and H-like (\ion{C}{vi}, \ion{N}{vii}, \ion{O}{viii}) lines \citep[see Table~2 and Fig.~3 of][]{detmers11}. The FWHM of non blended lines was about 4000\,\kms. 
For the triplets, neither the FWHM nor the individual-line fluxes could be disentangled. In particular, only for 
the resonance lines has a significant detection been found. We therefore took the FWHM of the resonance
line as a reference value and derived the upper limits of the intercombination and forbidden lines for both the \ion{O}{vii} and 
\ion{Ne}{ix} triplets. In Table\,~\ref{t:lines} we report the intrinsic line luminosities.\\

The luminosity of the \fek\ line (Table~\ref{t:lines}) has been measured by the \epi\ instrument. The line is formed by a constant,
narrow, component plus a broad, smoothly variable component \citep{ponti13}. We do not consider here the narrow component whose FWHM is not resolved by \xmm. This component
is not variable on long time scales and may be caused by reflection from regions distant from the black hole, like the molecular torus.  
The broad and variable component of the \fek\ line has a FWHM of about 15,000-30,000\,\kms\ which may probably partly arise from the BLR \citep{ponti13}. 
The \epi\ spectrum of \mrk\ also shows hints of highly ionized lines from \ion{Fe}{xxv}
and \ion{Fe}{xxvi}. These are too ionized to be produced in the BLR \citep[e.g.][]{costantini10}, but they are likely to come from a hot inner part of the molecular 
torus \citep{costantini10, ponti13}. Thus we do not include these lines in this analysis.

\subsection{The UV broad lines}\label{par:uvlines}
In the HST-COS modeling of the emission lines, more than one Gaussian component is necessary to fit the
data (see Table~3--6 and Fig.~4 of K11). The most prominent lines (i.e. \Lya\ and \civ) show as many as four distinct components. 
A first narrow component has a FWHM of about 300\,\kms, then an intermediate component with FHWM
of about 1000--3000\,\kms\ and a broad component with 4000--5000\,\kms\ are also present in the fit. Finally, a very broad component with FWHM
of about 9000--10\,000\,\kms\ is present for the most prominent lines (Table~\ref{t:lines}). We ignored in this study the narrow component (FWHM$\sim$300\,\kms), which is unlikely to
be produced in the BLR but should rather come from the Narrow Line Region (NLR). Due to the complex and extended morphology of the narrow-line emitting gas, the distance of the NLR 
in this source is uncertain \citep{phillips83, fischer15}. From the width of the narrow lines, 
the nominal virial distance ranges between 6 and 13\,pc, depending on the black hole mass estimate (see Sect.~\ref{par:geometry}). Note that with respect to Table~3 in K11, we summed the doublet luminosities as in many cases they are partially blended with each other. 
We corrected the line fluxes for the Galactic extinction (E(B-V)=0.057) following the extinction law in \citet{cardelli89}. The errors listed in Table~\ref{t:lines} are discussed below (Sect.~\ref{par:errors}).\\   
We also used the archival FUSE data,
which offer the flux measurements of shorter wavelength lines. The drawback of this approach is that the FUSE
observations were taken about 10 years before our campaign (in 1999--2000). In this time interval the source might
have changed significantly its flux and emission profiles. In this analysis we chose the 1999 observation \citep{kriss00}, 
as in that
occasion the flux was comparable to the HST-COS data in the overlapping band and the FWHM most resemble the present data. In Table~\ref{t:lines} we report the FUSE line
luminosities used in this paper. Also in this case we summed doublets and the blended lines.

\subsection{The optical broad lines}
The Optical Monitor (OM) data were collected at the same time as the X-ray data presented in this paper. The data reduction
and analysis has been presented by \citet{med11} and included correction for Galactic absorption and subtraction of the stellar contribution from the host galaxy from both the continuum and the emission lines. The optical grism data, covering the 3000--6000\,\AA\ wavelength range, 
displayed indeed clear emission lines of the Balmer series \citep[see Table~3 and Fig.~4 in][]{med11}. For 
the H$\alpha$ line two line components could be disentangled into a narrow and unresolved component with a flux of
$\sim3.3\times10^{-13}$\,erg\,cm$^{-2}$\,s$^{-1}$ and a broader component, with a FWHM of $\sim4300$\,\kms. For the other lines of
the series, the narrow component could not be disentangled. In order to obtain an estimate of the flux of the broad component alone,
we simply scaled the flux of the H$\alpha$-narrow component for the line ratio of the other lines of the Balmer series \citep[e.g. ][]{rafanelli}. We then subtracted the estimated narrow-line flux from the total flux measured by OM, resulting
in a relatively small correction with respect to the total line flux. The intrinsic luminosities of the broad lines 
are reported in Table\,\ref{t:lines}.    

\subsection{Notes on the uncertainties}\label{par:errors}


The uncertainties associated with the measurements are quite heterogeneous, reflecting the
different instruments' performances. 
In the UV data, the statistical errors on the fluxes are extremely small (2--4\%, K11). However, the line luminosities are affected by
additional systematic uncertainties, due to the derivation of specific line components, namely the ones coming from the BLR, among a blend of
different emission lines, with different broadening, fluxes, and velocity shifts. For instance, the \civ\ line doublet at
1548\,\AA\ is the sum of as much as seven components which suffer significant blending (K11). As seen in
Sect.~\ref{par:uvlines}, only for the strongest lines could three broad lines widths be disentangled. However, for the lower-flux
lines this decomposition could not be done, leaving room for additional uncertainties on the line flux of the broad
components. Therefore, we assigned more realistic error bars to the UV data. 
We associated an error of 20\% to the fluxes, which is roughly based on the ratio between the narrow and
the broad components of the \civ\ doublet. We also left out from the final modeling \ion{Si}{ii} ($\lambda\lambda 1260, 1526$\,\AA) and \ion{N}{iv}] as in the original COS data (see K11) those lines
are easily confused with the continuum and are therefore affected by a much larger, and difficult to quantify, uncertainty than the one provided by the statistical error. 

The line fluxes and widths observed by FUSE in 1999 may also be different from 2009. 
K11 estimated that the continuum flux was 34--55\% lower and the emission
lines could have been affected. In order to take into account the possible line variability, we assigned to the FUSE detections an error of 40\% on the flux. We also left out the \niii\ from the fit. 
Being \niii\ a weak and shallow line, only a very-broad component is reported, which may be contaminated by the continuum emission \citep{kriss00}.

For the \ovii\ triplet, in the RGS band, we summed up the values of the line fluxes, but formally retaining the percentage error on the best-measured line, as
upper limits were also present \citep{detmers11}. We considered \cvi\ and \nvii\ as upper limits because the detection was not more significant than 2.5$\sigma$ 
in the RGS analysis. However we used these two points as additional constraints to the fit, using them as an upper limit value on the model.\\
For the iron broad component, detected at 6.4\,keV in the \epi\ spectrum, which we consider in this work, we also summed the \fek\ line with the \fekb\ line (about 10\% of the flux of the \fek) as we do in the
model.

\section{The data modeling}\label{par:model}

\subsection{The LOC model}\label{par:loc_modeling}
In analogy with previous works \citep{costantini07, costantini10}, we interpret the broad emission features using a
global model. In the "locally optimally-emitting cloud" model, the emerging emission spectrum is not dominated by the details of the
clouds (or more generally gas layers), but rather on their global distribution in hydrogen number density, $n$, and radial distance from the source, $r$ \citep{baldwin}. The gas distribution is indeed
described by the integrated luminosity of every emission line, weighted by a powerlaw distribution for $n$ and $r$:\\
\begin{equation}
L_{\rm line}\propto\int\int L(r,n)\ r^{\gamma}\ n^{\beta}\ dr\ dn.
\end{equation} 

The powerlaw index of $n$ has been reported to be typically consistent with unity in the LOC analysis of quasars \citep{baldwin97}. A steeper (flatter) index for the density
distribution would enhance regions of the BLR where the density is too low (high) to efficiently produce lines \citep{korista00}. Here we assume the index $\beta$ to be
unity. Following C07, the density ranged between
$10^{8-12.5}$\,cm$^{-3}$. This is the range where the line emission is effective \citep{korista97a}. The radius ranged between $10^{14.75-18.5}$\,cm, to include also the possible X-ray emission from the lines, 
in addition to the UV and optical ones (C07).
The gas hydrogen column density was fixed to $10^{23}$\,cm$^{-2}$ where
most of the emission should occur \citep{korista97a,korista00}. Besides, the emission spectrum is not significantly 
sensitive to the column density in the range $10^{22-24}$\,cm$^{-2}$ \citep{korista00}. The grid of parameters has been constructed using Cloudy (ver.~10.00), 
last described in \citet{ferland13}. For each point of the grid, $L(r,n)$ is calculated and then integrated according to Eq.~1.

The emitted spectrum is dependent on the spectral energy distribution (SED) of the source. In this case we benefited from the simultaneous
instrument coverage from optical (with OM) to UV (with HST-COS) and X-rays (\epi\ and \inte). 
As a baseline we took the SED averaged over the 40-days \xmm\ monitoring campaign
\citep[labeled standard SED in Fig. 3 of][]{kaastra1}, taking care that the SED is truncated at infrared frequencies (no-IR case in that figure). 
Although the accretion disk must have some
longer-wavelength emission, most of the infrared part (especially the far-IR bump) is likely to emerge from outer parts of the system, like the molecular torus. An overestimate
of the infra-red radiation would mean to add free-free heating to the process. This effect becomes important at longer wavelengths as it is 
proportional to $n^2/\nu^2$, where $\nu$ is the photon frequency. Free-free heating significantly alters the line ratios of e.g. \Lya\ to \civ\
or \ovi\ \citep[][]{ferland02}. To avoid this effect, we truncated the SED at about 4\,$\mu$m. During the \xmm\ campaign the light 
curve of both the hard (2--10\,keV) and the soft (0.5--2\,keV) X-ray flux raised gradually up to a factor 1.3 and decreased of about the same factor in about one month \citep{kaastra1}. The OM photometric points
followed the same trend \citep[e.g. Fig.~1 of][]{med11}. Variations of the continuum fitting parameters \citep[discussed in ][]{med11,pop13} 
were not dramatic. Therefore at first order, the SED did not change significantly in shape, while varying in normalization.

\begin{figure}
\begin{center}
\resizebox{0.44\textwidth}{!}{\includegraphics[angle=0]{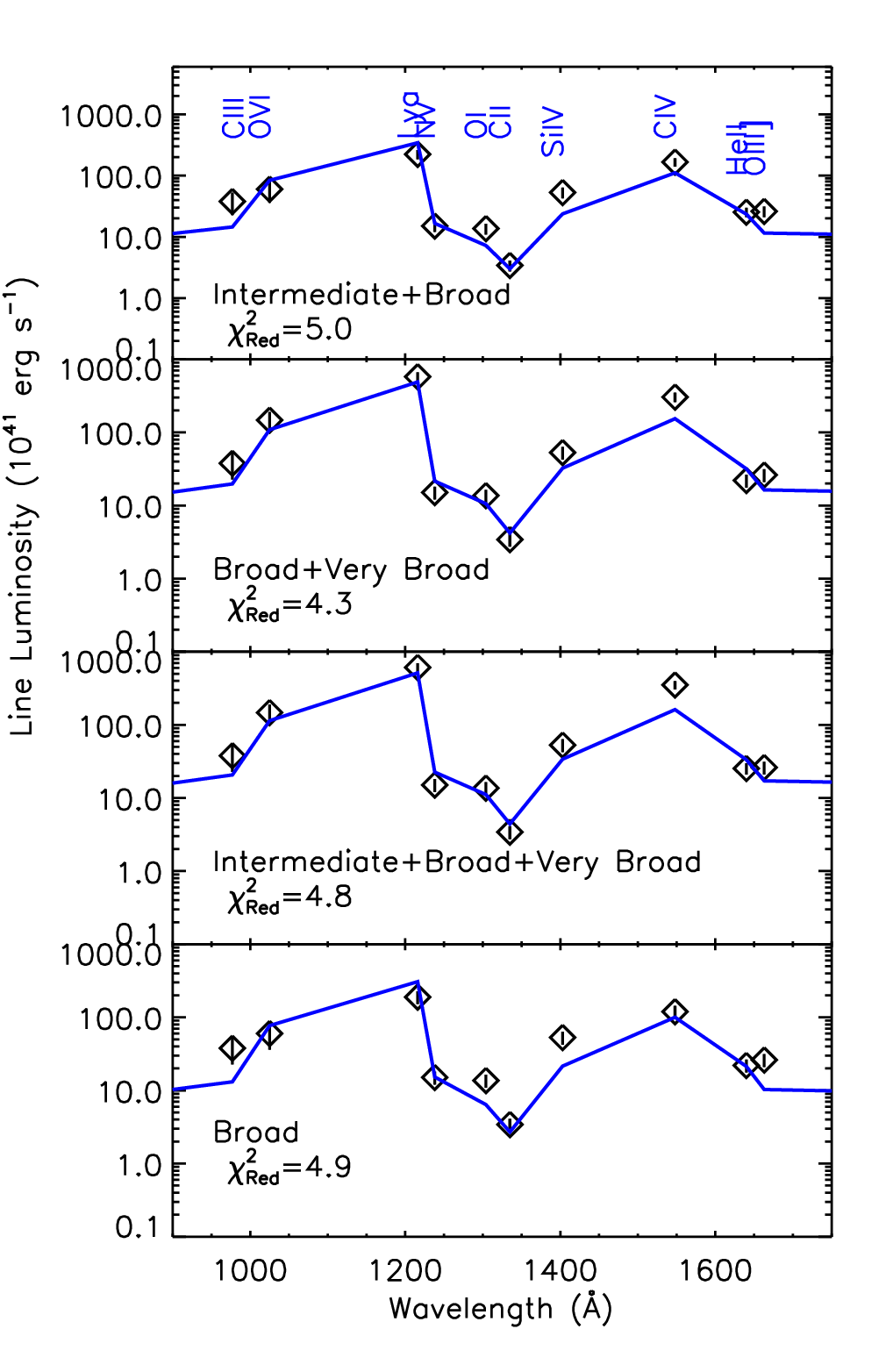}}
\end{center}
\caption{\label{f:bad_uvnarrow} The LOC fitting considering the sum of the broad and very-broad line-components provides consistently a better description of the data for any other choice of parameters. 
From top to bottom: combination of intermediate+broad, broad+very-broad, intermediate+broad lines+very-broad and broad lines alone. 
In this example, we only show the fit to the UV data.}
\end{figure}

\subsection{The LOC fitting}\label{par:fit}

The best-fit distribution of the gas in the black hole system is dependent on many parameters using the LOC model. The radial distribution and the covering factor of the gas, which are the
free parameters in the fit, in turn depend on pre-determined parameters, namely the SED, the metallicity (that we assume to be solar for the moment, see Sect.~\ref{par:abundances}), and the inner and outer radii of the gas. Moreover, broad lines
measured in an energy range covering more than three decades in energy, are likely to arise from gas distributed over a large region with inhomogeneous characteristics.

In fitting our model, we considered four different UV line-widths combinations, namely intermediate+broad, broad+very broad, intermediate+broad+very broad 
as well as broad lines alone (Table~\ref{t:lines}). We also selected six bands over which to perform the
$\chi^2$ test on the line flux modeling i.e. optical, X-rays, UV, optical+UV, X-rays+UV and X-rays+UV+optical. The individual bands are defined by the instruments used (Table~\ref{t:lines}). 
We used an array of six possible inner radii, ranging from log\,$r$=14.75 to 17.7\,cm (the actual outer radius being log\,$r$=18.5\,cm) to construct the model. Considering all combinations
of parameters, we obtain 144 different fitting runs. 
Whenever a limited number of lines (e.g. the UV band lines alone) are fitted, the model is extrapolated to the
adjacent bands to inspect the contribution of the best-fit model to the other lines. Not all the runs are of course sensitive to all the parameters. For instance a run which fits the
X-ray band only will be insensitive to any UV line widths. 
Free parameters of the fit are the covering factor $C_V$ of the gas and the slope $\gamma$ of the
radial distribution. The covering factor ($\Omega/4\pi$, where $\Omega$ is the opening angle) measured by the LOC is the fraction of the gas as seen by the source. The value of $C_V$ is constrained to lie in the range 0.05--0.6, based on the range of past estimates for the BLR obtained with different techniques (see Sect.~\ref{par:geometry}). 
In the following we describe the dependence of the fit on the different parameters, based on the goodness of fit. 

In Fig.~\ref{f:bad_uvnarrow} we show the comparison among best-fits with different line widths for the UV lines. Considering the same band (the UV only here), the inclusion of the
intermediate component (Sect.~\ref{par:uvlines}) systematically slightly worsen the fit.  
For simplicity, in the following we describe the sum of the broad and very broad components only, as they provide a slightly better fit, 
although the other combinations were also always checked in parallel.

\begin{figure}
\begin{center}
\resizebox{\hsize}{!}{\includegraphics[angle=90]{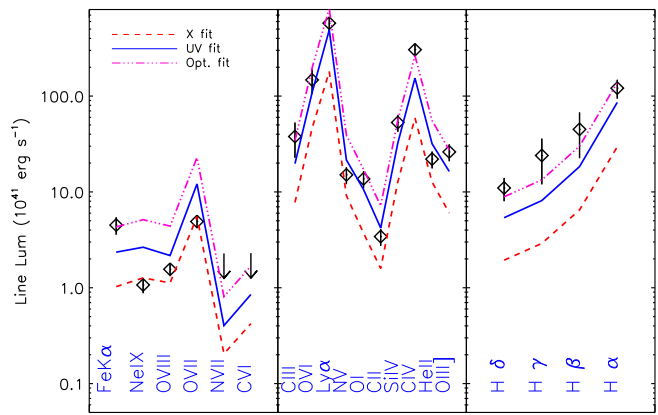}}
\end{center}
\caption{\label{f:bands} LOC fits over individual bands: X-rays (dashed line), UV (solid line) and optical band (dash-dotted line).}
\end{figure}

We show the fits in the different wavelength bands in Fig.~\ref{f:bands}. In Table~\ref{t:1comp} the best fit parameters are shown for 
each combination of bands, using the full range of radii. We note that the UV data certainly dominate
the fit, by virtue of the larger number of data points with a relatively smaller error bar. The global fit however does not completely explain the high- and low-energy ends of the data. A fit based on
the X-ray data under-predicts the UV and optical data, maybe suggesting the presence of an additional component. On the other hand, the fit based on the optical band, well describes both
the optical and the \ion{C}{iv} UV line, albeit with a large overestimate of the rest of the UV and X-ray lines.

\begin{figure}
\begin{center}
\resizebox{\hsize}{!}{\includegraphics[angle=90]{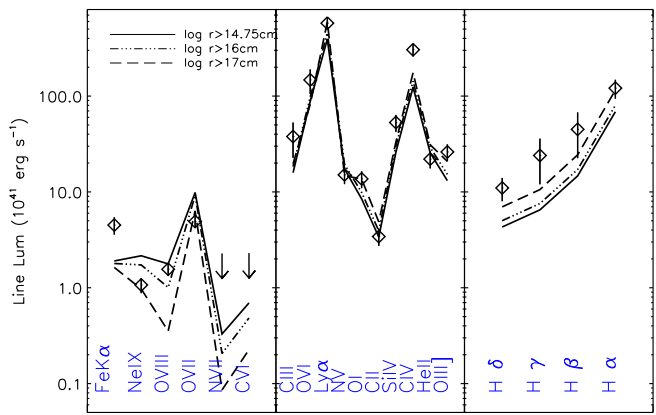}}
\end{center}
\caption{\label{f:radius} LOC fits over the whole spectral band (X-UV-optical) with three representative inner radii. The X-ray band is best fitted for smaller inner radii, while the Optical band
may be better described if larger radii are used.}
\end{figure}

The LOC fitting depends also on the inner radius over which the radial gas distribution is calculated. We fitted the data for a choice of six inner radii, roughly separated by 
half a decade. In Fig.~\ref{f:radius} a fit considering all the data is plotted for a selection of inner
radii. We see that while the UV band is only marginally affected by the inner radius choice, this parameter can make a difference for the optical and X-ray bands. 

The application of a a single component of the LOC model with some tunable parameters, does not totally explain the data. In the following we explore other effects that may play a role in
the line emission. 

\begin{table}
\caption{\label{t:1comp} Results the LOC fitting considering different spectral bands.}
\begin{center}
\begin{tabular}{llll}
\hline
\hline
& $\gamma$ & $C_V$ & $\chi^2_{\rm Red}$\,({\it dof})\\
\hline
O 		& $1.10\pm0.5$ & $>0.6$ & 0.9 (2)\\
UV		& $1.05\pm0.08$ & $<0.05$ & 4.3 (8)\\ 
X		& $1.13\pm0.23$ & $>0.6$ & 1.5 (4)\\
UV+X	& $1.05\pm0.05$ & $<0.05$ & 4.9 (14)\\
UV+O	& $1.0\pm0.1$ & $<0.05$ & 3.4 (12)\\
UV+O+X	& $1.05\pm0.06$ & $<0.05$ & 4.4 (18)\\
\hline
\end{tabular}
\end{center}
Notes: $\gamma$ is the slope of the radial distribution of the line luminosities. $C_V$ is the covering factor of the
BLR gas. $\chi^2_{\rm Red}$ is the reduced $\chi^2$ and $dof$ are the degrees of freedom.

\end{table}

\subsection{Extinction in the BLR}\label{par:ext_blr}
As seen above, the application of the LOC model to \mrk\ points out that the optical lines are systematically underestimated. 
A possible solution is to include extinction in the BLR itself. The UV/optical
continuum \mrk\ is not significantly reddened \citep[][]{osterbrock77}. However, the dust may be associated only to the line emission region, in a way that the continuum that we measure would be unaffected
\citep[][]{osterbrock06}. In principle, the \heii(1640\AA)/\heii(4686\AA) ratio would be an indicator of reddening intrinsic to the BLR \citep[e.g.][]{osterbrock06}. In practice, both lines are severely
blended with neighboring lines and with the wing of higher flux lines, namely \civ\ for \heii(1640\AA) and \Hbeta\ for \heii(4686\AA) \citep{bottorff02}. In our case, the observed line ratio is very low
($\sim1.7$) when compared to the theoretical value derived from the LOC model \citep[6--8;][]{bottorff02}. This line ratio would imply an extinction E(B-V)=0.18 \citep[eq.~4 of][]{annibali10}, 
when a Small Magellanic Cloud (SMC) extinction curve, possibly more appropriate for AGN, is used \citep[][]{hopkins04,willot05}. Knowing the uncertainties (i.e. line blending) 
associated to our \heii\ measurements, we took this value as the upper limit of a series of E(B-V) values to be applied to our lines. Namely, we tested E(B-V)=(0.18, 0.15, 0.10, 0.075, 0.05, 0.025). 
We then corrected accordingly the observed optical and UV fluxes
\citep[][]{annibali10}. For the lines observed by FUSE (Table~\ref{t:lines}), we extrapolated the known SMC extinction curve with a $\lambda^{4}$ function to reach those wavelengths. 
The extinction in the X-rays has been simply estimated using the E(B-V)-$N_{\rm H}$ relation provided in \citet[][]{predehl95}, considering a SMC-like selective to total extinction ratio $R_V$
of 2.93 \citep{pei92}.\\
When only the UV lines are modeled, the $\chi^2$ method chooses lower values of the BLR extinction, with a final $\chi^2_{\rm red}$ of 6.1 ({\it dof}=8) for E(B-V)=0.025.
The effect of the BLR extinction is relatively modest in the X-ray band and mainly affecting the \ovii\ lines. 
However, any value of the BLR extinction largely overcorrects the Balmer series lines. Therefore when also the optical lines are included in the model, the resulting fit
becomes even worse ($\chi^{2}_{red}=16-20$, for 18 {\it dof}).

\subsection{A two-component LOC model}\label{par:two}
     
\begin{figure}
\begin{center}
\resizebox{\hsize}{!}{\includegraphics[angle=90]{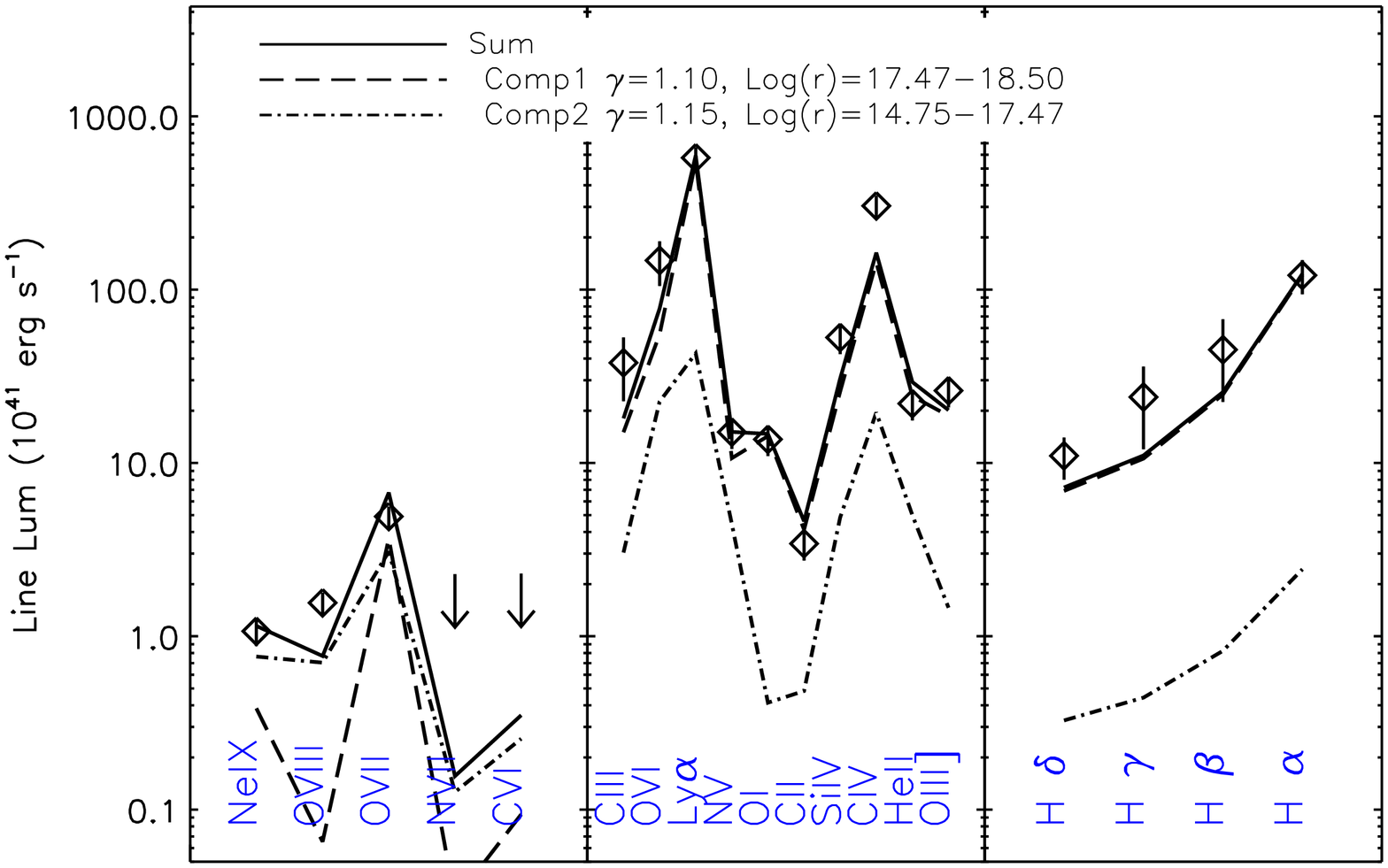}}
\resizebox{\hsize}{!}{\includegraphics[angle=90]{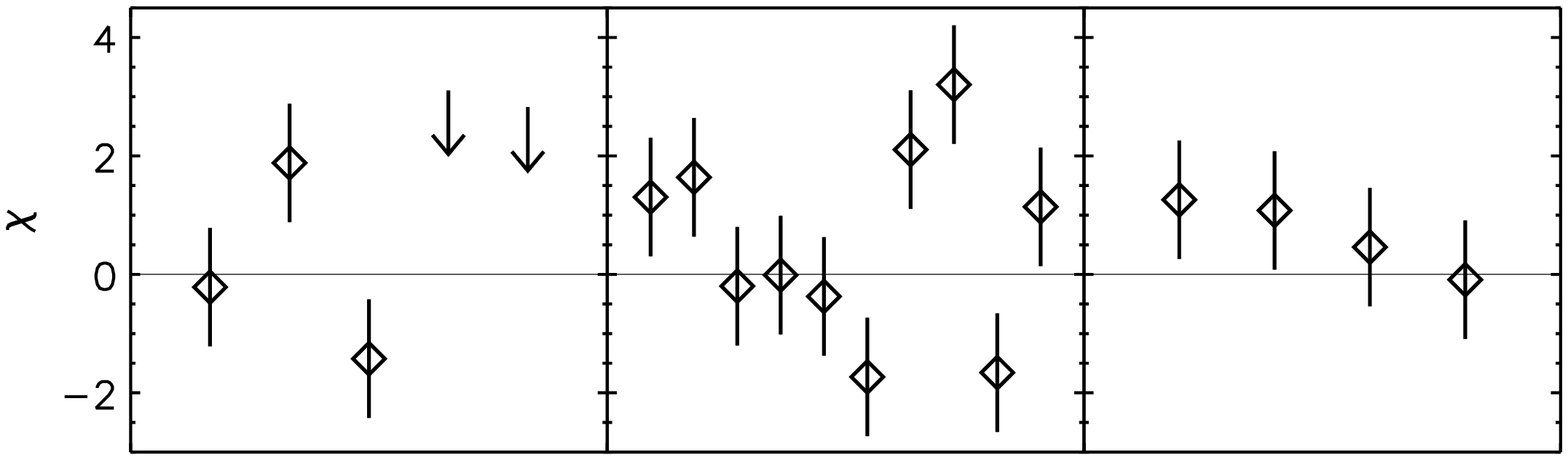}}
\end{center}
\caption{\label{f:regions} Upper panel: LOC fit with two components, acting in different regions near the AGN (Model~3 in Table~3). 
X-ray data are best fitted by a component near the black hole, while the optical data are better 
fitted by a further-away component. Lower panel: residuals to the fit.}
\end{figure}

A single LOC-component does not provide a fully satisfactory fit. This is not surprising, given the large range of ionization potentials of the lines. 
Therefore we attempt here to test a two-component model. 
As before, for each of the two components we fit all the combinations of line widths (as in Fig.~\ref{f:bad_uvnarrow}). We first considered the whole range of radii (Model~1 of Table~\ref{t:total}). Then we made the inner radius (as defined in
Sect.~\ref{par:fit}) of both components vary (Model~2 in Table~\ref{t:total}). Finally, we took into account the different emissivity depending on the size of the region, by varying also the outer radius. To do 
this, we divided the radial range into four regions (starting at log$r=14.75, 15.53, 16.56, 17.47$), 
in order to have roughly an order of magnitude difference between two adjacent radii. 
We then considered for each component all combination of adjacent regions (or single regions). 
Therefore we have a total of 10 options for the size of each component of the gas (Model~3 in Table~\ref{t:total}). 
Note that for each run the inner and outer radii were fixed parameters. 
We fitted the whole band (X-ray, UV, optical: XUVO) for the two LOC components. The fit is driven by the UV band, where the uncertainties on the data are the smallest.
We note that the slope of the powerlaw is dependent on the covering factor, as flatter slopes ($\gamma<1.1$) systematically correspond
to very small covering factors ($C_V<0.05$). Conversely, the upper limit we set for the covering factor ($C_V$=0.6) corresponds to steeper radial slopes \citep[see also][]{korista00}. 
The covering factor has the effect of regulating the predicted line luminosities. A steeper radial distribution would enhance the lines at smaller radii, 
where the gas illumination is stronger. Therefore a larger $C_V$ would be required to tune down the line luminosities. 
On the contrary, a flatter slope, would lower the contribution of the strong-illuminated region, while the outer radii are enhanced. 
However, the radiation field lowers with the distance, therefore a smaller $C_V$ is necessary to adapt the predicted fluxes to the real data. 
In the last line of Table~\ref{t:total} we report the reduced $\chi^2$. The reduced $\chi^2$ never falls below $\sim$2, even for the better-fitting models. 
This is especially due to the outlying data points, namely \siiv, \civ,
 and \heii. The exclusion of the \fek\ was not resolutive, as this line has a larger uncertainty with respect to the UV lines.   
Fig.~\ref{f:regions} refers to Model 3. As expected, the more sensitive lines were the optical and the
X-rays, respectively. A highly ionized component, extending down to
log$r<$14.7\,cm is necessary in order to reproduce the \oviii\ and \neix\ lines in the X-rays. All the optical lines and part of the \civ\ line 
are best fitted by adding a component with a larger inner radius. 

\begin{table*}
\caption{\label{t:total} Best fit parameters for a two-component model and different emitting regions.}
\begin{center}
\begin{tabular}{llll}
\hline\hline
& {\bf Model 1} & {\bf Model  2} & {\bf Model 3}\\
\hline
{\bf Comp 1}&&&\\
$r_{in}$ & 14.75 & 17.0 & 17.47\\
$r_{out}$ & 18.5 & 18.5 & 18.5\\
$\gamma$ & $1.59\pm0.03$ & $1.04\pm0.02$ & $1.10\pm0.02$\\
$C_V$ & $<0.05$ & $<0.05$ & $>0.6$\\
\hline
{\bf Comp 2}&&&\\
$r_{in}$ & 14.75 & 14.75 &14.75\\
$r_{out}$ & 18.5 & 18.5 & 17.47\\
$\gamma$ & $1.06\pm0.02$ & $1.17\pm0.02$ & $1.15\pm0.02$ \\
$C_V$ & $<0.05$ & $>0.6$ & $>0.6$\\
\hline
$\chi^2_{red}$\,({\it dof}) & 5.2(15)&  2.5(15)& 2.3(15)\\
\hline
\end{tabular}
\end{center}
Notes:\\
The parameters are: $r_{in}$, the inner radius; $r_{out}$ the outer radius; $\gamma$, 
the slope of the radial distribution and $C_V$, the covering factor. Note that the \fek\ line is excluded from this fit. 
model~1: the emissivity occurs over all radii for both components.\\
model~2: the two components have different inner radii.\\
model~3: both the inner and outer radii of the emissivity is variable for both components.\\
\end{table*}

\section{Discussion}\label{par:discussion}

\subsection{Abundances and the influence of the SED}\label{par:abundances}
Abundances in the BLR should be either solar or super-solar. The metal enrichment should come from episodic starburst
activity \citep{romano02}. The \nv\ line is often taken as an abundance indicator in AGN since it is a product of the 
CNO cycle in massive stars. 
Using the broad component ratios in our data for \nv/\civ\ and \nv/\heii, the diagnostic plots of \citet{hamann02} 
suggest abundances in Mrk 509 of $1<Z<3$ \citep[see][for the limitations in determining abundances in the BLR]{steen11}. In this analysis we
considered a SED with solar abundances, as defined in Cloudy. We therefore also tested the fits presented above using a metallicity 3 times solar. 
The fits obtained are systematically worse ($\Delta\chi^2=2-7$ for the same number of degrees of freedom). This suggests that the abundances are close to solar.

The present HST-COS data were taken 20 days after the last \xmm\ pointing \citep{kaastra1}, as the closing measurements of the campaign, which lasted in total about 100 days. 
Spectral coverage simultaneous to HST-COS was provided instead by both \ch-LETGS \citep{ebrero11} and Swift-XRT \citep{med11}. We used the average SED recorded, 20--60 days before the HST-COS observation, by the \xmm\ instruments. The choice of SED is very important in the BLR modeling, as different lines
respond on different time scales to the continuum variations \citep{korista00,peterson04}. Reverberation mapping studies of \mrk\ report that the delay of the  
\Hbeta\ with respect to the continuum is very long \citep[about 80 days for \Hbeta,][]{carone96, peterson04}. However, higher ionization lines respond faster to
the continuum variations. Taking as a reference the average \Hbeta/\civ\ delay ratio for NGC~5548 \citep{peterson04}, for which, contrary to \mrk, a large set of line measurements is available, we obtain that 
the \civ\ line in \mrk\ should roughly respond in 40 days. A similar (but shorter) time delay should apply to the \Lya\ line \citep{korista00}. 
This delay falls in the time interval covered by the \xmm\ data. Therefore our choice of SED should be appropriate for the modeling of at least the main UV lines. Variability of the X-ray broad
lines has been reported on years-long time scales \citep{costantini10}, however no short-term studies are available. We expect that the X-ray broad lines should respond promptly to the continuum variations, as they may be located up to
three times closer to the black hole with respect to the UV lines (C07). During the \xmm\ campaign the flux changed at most 30\%, with a minimal change in spectral shape 
(Sect.~\ref{par:loc_modeling}). The used SED should therefore represent what the BLR gas see for the X-ray band.  
However, for the optical lines the used SED might be too luminous as the we observed an increase in luminosity by about 30\% during the XMM-Newton campaign, 
and as seen above, the time-delay of the optical lines may be large.

\subsection{The UV-optical emitting region}
\begin{figure}
\hspace{-0.8cm}
\hbox{
{\includegraphics[angle=90,height=4cm,width=5cm]{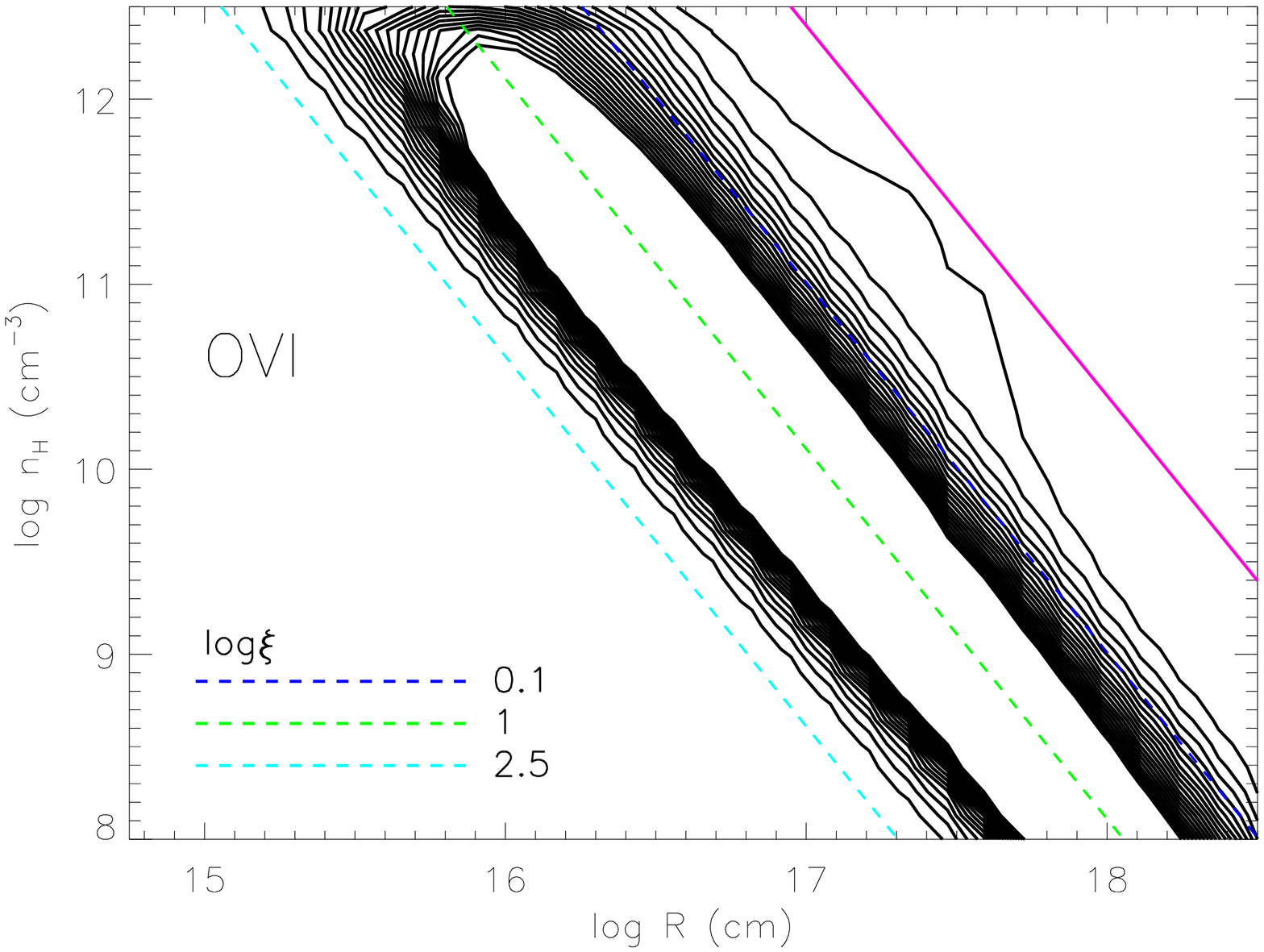}}
\hspace{-0.2cm}
{\includegraphics[angle=90,height=4cm,width=5cm]{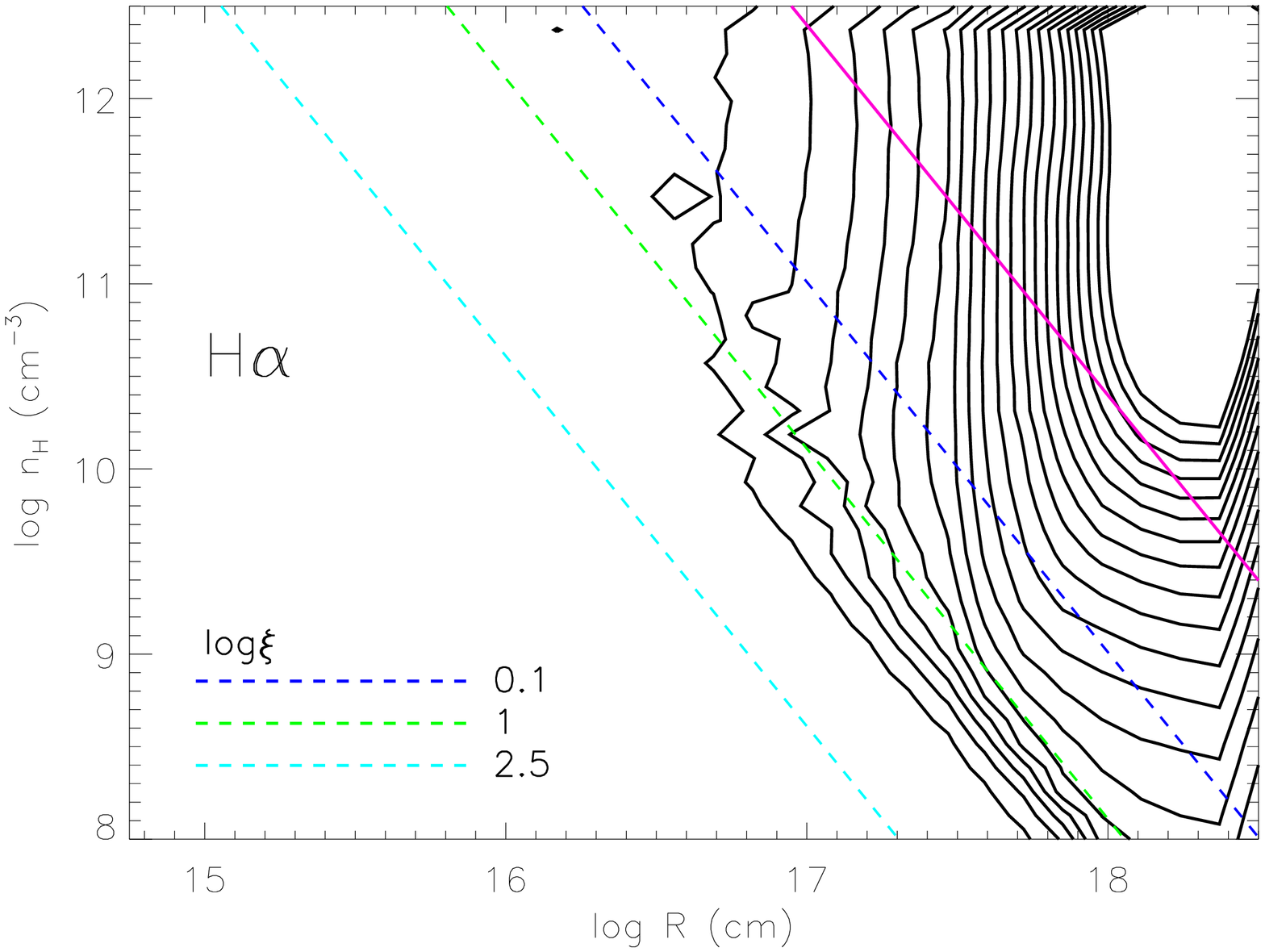}}
}

\caption{\label{f:contour} contour profiles of \ovi\ and \Halpha\ as a function of density and distance, here using a radial slope $\gamma$ of 0.10. The dashed lines indicates constant
ionization parameters, as detailed in the legend. The solid magenta line follows the density prediction of the pressure confined emission model of \citet{baskin14}.}
\end{figure}

The LOC model has been extensively used to model the UV and optical lines of AGN \citep[e.g.][]{korista97a}. 
In this study we find that a single radial distribution of the gas over the whole range of radii, applied to the UV band, would have a slope $\gamma\sim1$, as
prescribed by the standard LOC model (Table~\ref{t:1comp}). The covering factor is unconstrained as it hits the lower limit that we imposed on this parameter. 
As in the case of Mrk~279 (C07), the \civ\ line is a systematic outlier. This line may obey to mechanisms other than pure gravitation
(e.g. inflows/outflows) or may arise in a geometrically different region than e.g. the optical lines \citep[e.g.][ and references therein]{goad12}. Finally, \Lya\ and \cvi\ are found in some sources to
respond on a slightly different time scale to the continuum variation. In the case of NGC~5548 this difference in response is of the order of 20 days \citep[][ Sect.~\ref{par:abundances}]{korista00}. This may account for some of the
mismatch between the two lines in our fit. 
As tested above (Sect.~\ref{par:ext_blr}), extinction in the BLR of \mrk\ must be negligible, therefore the discrepancy with the model cannot be ascribed to dust
in the emitting region. The ionization of the BLR follows the rules of photoionization. In particular for a given UV-emitting ion \citep[e.g. \civ, \Lya, \ovi, as detailed in ][]{korista97a}, 
the ionization parameter remains constant throughout the region (dashed lines in Fig.~\ref{f:contour}). 
Note that for lower ionization lines (namely, the Balmer
lines, Fig.~\ref{f:contour}, right panel), density effects come into play besides pure recombination \citep{kk79,osterbrock06} and the ionization parameter does not follow the emission contour
\citep{korista97a}.
This model does not require a universal ionization parameter, because of the assumption of the stratified nature of the gas. A pressure confined gas model, which may also allow for a
range of ionization parameters in a stratified medium, would also predict, given a bolometric luminosity, a gas hydrogen density as a function of radius \citep[eq 21 in][]{baskin14}. This prediction is
drawn in Fig.~\ref{f:contour} (magenta solid line), using $L_{bol}\sim3L_{1350\,\AA}$ \citep{kaspi05}, where $L_{1350\,\AA}$ has been extrapolated from the average SED of Mrk~509
\citep{kaastra1}. This density prediction is not too far off, however it overestimates the optimal emitting region density 
of the higher ionization ions (an example is given in the left panel of Fig.~\ref{f:contour}), while it would match the Balmer lines emitting region (right panel).

\subsection{The size and gas distribution of the BLR}\label{par:geometry}

Several arguments point to a natural outer boundary for the BLR which should be intuitively given by the dust sublimation radius \citep[][]{suganuma06,landt14}. 
For \mrk, this radius corresponds to $3.6\times10^{18}$\,cm \citep{mn12}. 

The maximum radius of our LOC model is $3\times10^{18}$\,cm. An expansion of the BLR outer radius to $7.6\times10^{18}$\,cm does not improve the fit. This is a natural
consequence of the LOC model construction. For radial distributions with slopes $\gamma\gtsim 1$ the line emissivity of
some major lines (\ovi, \civ) already drops at $10^{18}$\,cm \citep[C07,][]{baskin14}. 
Therefore our fit is consistent with a confined BLR region, possibly within the sublimation radius. 

The radius of the BLR has been found to scale with the UV luminosity. If we take the \civ\ line as a reference, $R_{\rm
BLR}=2\times10^{16}h_0L_{42}^{0.5}$(\civ), where $h_0$ is the Hubble constant in units of 100\,\kms\ and $L_{42}$ is the \civ\ luminosity in units of $10^{42}$\,erg\,s$^{-1}$ 
\citep{peterson97}. For \mrk, the radius of the BLR based on this equation is $\sim2.6\times10^{17}$\,cm. Using instead the known relation between the size of the \Hbeta\ emitting region
and the luminosity at 5100\AA, we obtain, for \mrk, $R_{\Hbeta}\sim1.2\times10^{17}$\,cm \citep{bentz13}.\\
In our fit the location of the UV emitting lines is consistent with these estimates, as, although UV lines are efficiently emitted in a large range of radii
(Fig.~\ref{f:radius} and C07), a large fraction of the UV line luminosity could come from radii $\geq 10^{17}$\,cm (Model 2,3 in Table~\ref{t:total}, Fig.~\ref{f:regions}). 
Assuming Keplerian motion, the FWHM of our lines imply that the very-broad lines (FWHM$\sim$9000-10,000\,\kms) are located at approximatively $2.5-5\times10^{16}$\,cm, depending on the mass
of the black hole: $1.43\times10^{8}$\,M$_{\odot}$ \citep{peterson04} or $3\times10^{8}$\,M$_{\odot}$ \citep{med11}. For the broad lines (FWHM$\sim$4000-5000\,\kms) the distance would then be 
$1.3-2.5\times10^{17}$\,cm, consistent with our results for the UV-optical component. Finally for the intermediate lines (FWHM$\sim$1000-3000\,\kms) the calculated distance is $2-4\times10^{18}$\,cm.
The location of the line emitting gas is stratified, therefore these single-radius estimates are only taken as a reference. The very-broad and the broad lines are well within the estimated radius for
the BLR. The so-called intermediate line region could possibly bridge the BLR and the NLR \citep{baldwin97}.

In interpreting the BLR emission, we tested a two-component model, characterized not only by different radial distributions and covering factors, 
but also by different physical sizes and inner/outer radii of the emitting plasmas.

Our fits are not completely satisfying as important outliers, like \civ, are present. 
However, the best fit points to the interesting possibility that the optical and part of the UV region originates at larger radii 
(starting at $3\times10^{17}$\,cm), while the X-ray- and 
some fraction of the UV-emission regions would have an inner radius smaller than $6\times10^{14}$\,cm (as also found in C07) and a larger extension up 
to about the beginning of the optical BLR (Sect.~\ref{par:two}). 
This would point to a scenario in which the optical lines, including the \Hbeta, would come from the outer region of the BLR. 
Such a span in distance between the optical and the X-ray lines, would also imply for the latter a faster response-time to any continuum variation. 
Such an effect has not been systematically studied, although strong flux variation of the \ovii\ broad line has been observed before \citep{costantini10}. 
The inability to find a good fit with the present model, 
which assumes a simple plane parallel geometry, could suggest a more complex geometry.
Recently for instance an inflated geometry \citep["bowl geometry",][]{goad12} for the outer region, possibly confined by a dusty torus has been 
suggested using different approaches \citep{goad12,pancoast12,gg13}.

The covering factor was set in our fits to be in the range 0.05--0.6. The lower limit has been chosen following 
early studies on the relation between the equivalent width of the \Lya\ and the
covering
factor \citep[0.05--0.1, e.g.][]{cf88}. However subsequent studies, using among others also the LOC model technique, 
have pointed out that the covering factor can be larger: from 0.30 \citep[e.g.][and references therein]{maiolino01} up to 0.5--0.6 \citep{korista00}. 
The covering factor here is the fraction of the gas as seen by the source. This is equal to the observer's line of sight covering 
factor only if a spherical distribution of the gas is assumed. 
A more flattened geometry would then
reconcile a large covering factor with the fact that absorption lines from the broad line region are in general not observed in the optical-UV band. 
In our fits the covering factor is unconstrained. However large covering factors have been preferentially found when a 
two-component model was applied, especially when the inner and outer radius were allowed to vary for both components. 
The measured high covering fraction, necessary to explain the line luminosities of the two components, would then point to a gas with non-spherical geometry. 
As these two components are along our line of sight, they may be one
below the other, therefore the sum of the two $C_V$ can well be above one, as long as the individual covering factors do not cover entirely the source (i.e. $C_V<1$).        

Despite the extensive exploration of the impact of different parameters to the modeling, our analysis also underlines that a simple 
parameterization may be inadequate to explain the complexity of
the BLR. Reasons for not reaching a better fit include minor effects like possible different responses 
of \Lya\ and \civ\ to continuum variations, non-simultaneity of the FUSE data, and inhomogeneous
information on the broad-band line profiles. The $C_V$ may not be a simple step function, but the clouds/gas-layers may experience a 
differential covering factor for instance as a function of the distance or line ionization.  
A major effect would be the complex dynamics and geometry of the BLR, which needs more sophisticated models to be explained. 

\subsection{The iron line at 6.4 keV}
In this paper we include the 6.4\,keV \fek\ line, observed simultaneously with the other soft X-ray, UV and optical lines. 
The narrow and non-variable component, probably produced in distant regions, was not considered in the fit. We find that the derived emission of the BLR contribution to the broad \fek\ line
component is around 30\%, if we used a two-component model. The emission would happen at a range of distances from the source, although at 
small radii (log$r\gtsim$14.75\,cm) the emission is enhanced (Fig.~\ref{f:radius}). 
Note that fortuitously, a single component fit, based on the optical lines, would provide a perfect fit to the \fek\ line
(Fig.~\ref{f:bands}). However, such a gas would produce both UV and soft-X-ray line fluxes at least a factor of 6 larger than observed. 
A modest contribution ($\sim17\%$) of the BLR to the iron line has been also reported in Mrk~279, using non simultaneous UV and X-ray data \citep[][]{costantini10}.

\section{Conclusions}\label{par:conclusion}
In this paper we attempted to find a global explanation of the structure of the gas emitting broad lines in \mrk, from the he optical to the X-ray band using a simple parametrization of the BLR. 
This study is possible thanks to the simultaneous and long observations of \xmm\ and HST-COS. 

We find that lines broader than FWHM$>$4000\,\kms\ contribute to the bulk of the BLR emission. 
A two-component LOC model provides a statistically better, but not conclusive, description of the data. 
The two components are characterized by 
similar radial emissivity distribution ($\gamma\sim1.10-1.15$), but different size and distance from the central source. 
The X-rays and part of the UV radiation come from an inner and extended region
($r\sim5\times10^{14}-3\times10^{17}$\,cm), while the optical and part of the UV gas would be located at the outskirts of the BLR ($r\sim3\times10^{17}-3\times10^{18}$\,cm). This picture 
appears to be in agreement with recent results
on the geometry of the BLR, locating the \Hbeta\ line away from the ionizing source. However, more sophisticated parameterizations are needed to have a definitive answer.

The \fek\ broader line cannot completely be accounted for by emission from the BLR gas. The contribution of the BLR is around 30\% for this line.

\begin{acknowledgements}
 The Netherlands Institute for Space Research is supported
financially by NWO, the Netherlands Organization for Scientific Research. \xmm\ is an ESA science missions with instruments and contributions directly funded by ESA 
Members States and the USA (NASA). We thank the referee, E.~Behar for his useful comments. We also thank L. di Gesu for commenting on the manuscript and G. Ferland and F. Annibali for discussion on
extinction in the BLR and host galaxy. GP acknowledges support of the Bundsministerium f\"ur Wirtschaft und Technologie/Deutsches Zentrum f\"ur Luft-und Raumfahrt (BMWI/DLR, FKZ 50
OR 1408). P.-O.P. and SB acknowledge financial support from the CNES and franco-italian CNRS/INAF PICS. G.K. was supported by NASA through grants for
HST program number 12022 from the Space Telescope Science Institute, which is operated by the Association of Universities for Research in Astronomy, Incorporated, under NASA contract NAS5-26555.
\end{acknowledgements}

\end{document}